\documentclass[twocolumn,runningheads,natbib]{svjour2}
\smartqed  
\usepackage{graphicx}
\journalname{Astrophysics and Space Science}
\begin{document}

\title{Two decades of pulsar timing of Vela
}
\subtitle{}


\author{Richard Dodson \and
        Dion Lewis     \and Peter McCulloch
}


\institute{R. Dodson, Marie Curie Fellow \at
Observatorio Astron\'omico Nacional, Espa\~na\\
           \email{r.dodson@oan.es}
           \and
           D. Lewis \at
           CSIRO, Australia   \\
           \email{dion.lewis@csiro.au}
           \and
           P. McCulloch \at
           University of Tasmania, Australia
}

\date{Received: date / Accepted: date}

\maketitle

\begin{abstract}

Pulsar timing at the Mt Pleasant observatory has focused on Vela,
which can be tracked for 18 hours of the day. These nearly continuous
timing records extend over 24 years allowing a greater insight into
details of timing noise, micro glitches and other more exotic
effects. In particular we report the glitch parameters of the 2004
event, along with the reconfirmation that the spin up for the Vela
pulsar occurs instantaneously to the accuracy of the data. This places
a lower limit of about 30 seconds for the acceleration of the pulsar
to the new rotational frequency.
We also confirm of the low braking index for Vela, and the continued
fall in the DM for this pulsar.

\end{abstract}

\section{Introduction}
\label{sec:intro}

Mount Pleasant observatory, just outside Hobart in Tasmania,
Australia, has a 14-metre dish that has been dedicated to tracking the
Vela pulsar for two decades. This telescope is able to monitor the
pulsar for eighteen hours every day, and therefore has caught many
glitches `in the act'. As a crosscheck the older, but glitching, PSR
1644-4559 is observed for the six hours when Vela is set.  There is no
comparative dataset, and the conclusions we draw puts extremely tight
conditions on the pulsar Equation of State (EOS) by placing a number
of constraints on the models. An example of these would be that if the
spin-up is very fast the crust has to have a low moment of inertia,
therefore be very thin, and the coupling between the crust and the
interior super-fluid has to be strong (see, for example, discussion of
these issues in Epstein \& Baym (1992) and Bildsten \& Epstein (1989)).
%

Three uncooled receivers are mounted at the prime focus of the
14-metre to allow continuous dispersion measure determination.  Two
are stacked disk, dual polarisation with central frequencies of
635~MHz and 990~MHz additional to a right hand circular helix at
1391~MHz.  Bandwidths are 250~kHz, 800~kHz and 2~MHz respectively,
limiting pulse broadening from interstellar dispersion to less than
1\%.  The output is folded for two minutes giving an integrated pulse
profile of 1344~pulses.  The backend to the 990~MHz receiver also has
incoherent dedispersion across 8 adjacent channels allowing a study of
individual pulses. Results from these systems have been reported,
respectively, in McCulloch \etal\ (1998) and Dodson \etal\ (2002).

A new system, based on the PC-EVN VLBI interfaces Dodson \etal\
(2004). 
can produce TOA's with accuracy of the order of 0.1 msec every second
(as opposed to every 10 seconds with the single pulse or 120 seconds
with the multi-frequency systems). This interface is adapted from the
Mets\"ahovi Radio Observatories linux-based DMA, data collector, card
designed for VLBI digital inputs.
The two 40~MHz IFs from the 635~MHz feed provide the two
polarisations, and the data are recorded in a continuous loop two hours
long. This is halted by the incoherent dedispersion glitch monitoring
program. Coherent dedispersion is performed off-line, for the data
segment covering the glitch.

\section{The Glitch in 2004}
\label{sec:2004}

The glitch of 2004 occurred while the telescope was recording
data. Unfortunately the coherent de-dispersion system was not running
at that moment and the results obtained are more or less a repeat of
those in 2000. The instantaneous fractional glitch size was
$2.08\times10^{-6}$. This is the sum of the permanent and the decaying
terms. The full details are reported in Table 1, after
fitting with TEMPO (Taylor \etal\ 1970). A similar fast decaying term as
reported in the 2000 glitch can be seen in the data after the usual
model is subtracted, see Figure 
1, but it is only a few sigma above the noise.  This usual model
consists of permanent glitch components in the frequency and frequency
derivate, and other components which are co-temporal jumps in frequency
which decay away. A number of these are required to fit the data and
the decay timescales are denoted with $\tau_n$. Three decaying terms
have been known for sometime and these make up the usual model. In
Dodson \etal\ (2002)
a forth short term component was identified.
For a fuller discussion see that paper. In Figure 1 the longer
timescale terms are subtracted, and the residuals are plotted scaled
against the RMS. Time zero is the intercept of the post-glitch model
with the pre-glitch model, i.e. assuming an instant spin-up. The
indication of a spin-up would be negative residuals, of which there is
no sign. The positive residuals seen are modelled as a later glitch
epoch and a very fast decaying term.
Figure 
1a shows the glitch of 2000, where the signal was
clearly above the noise,
Figure 
1b shows the glitch of 2004, where the signal is
barely above the noise.
We present it as supporting evidence for the similar signal seen in
the 2000 data, but we are unable to draw more detailed conclusions
from such weak data.

\section{Observations over the last twenty years}
\label{sec:20}

The 14-metre telescope has recorded single channel two minute folded
data from July 10 1981 to October 1 2005, spanning 8857 days. It has
recorded incoherently averaged ten second folded data from 1998. It
observed on the day of a glitch for all ten events in that twenty year
period. After upgrading to a full Az-Alt telescope in 1987 it was able
to track the pulsar for 18 hours a day, and was therefore able to
catch the very moment of the glitch in 1988, 1991, 2000 and 2004, as
well as the first of the two in 1994. The two in 2000 and 2004 were
with the incoherent single pulse system which is folded over ten
seconds to give a good signal to noise. In all of these cases there is
{\bf no detectable spin up}. Figure
2 shows a montage of the recorded glitch events. Compare this to the
stately, half day, spin up of the crab pulsar 
(Wong \etal\ 2001).
The upper limit from the 2000 and 2004 de-dispersed data is that the
spin-up occurs in less than 30 seconds. It is most unfortunate that,
as yet, we have no observations with the coherent de-dispersion
system, as these very fast spin up times measure directly the moment
of inertia, and therefore the thickness, of the Vela pulsar crust
(Epstein \& Baym, 1992). Figure
3 shows the fitted $\dot{{\rm F}}$ over the twenty years.
As an independent check of the braking index calculation of Lyne
\etal\ (1996)
we applied their method (used on the JPL data from 1969
to 1994, nine glitches) to our data (1981 to 2005, ten glitches). Here
they assume that some months after the glitch the short term
relaxations have decayed away, and the build-up of timing noise has not
yet contaminated the rotation rate. Therefore by fitting $\ddot{{\rm F}}$
to the slope $\dot{{\rm F_{150}}}$, the deceleration 150 days after
the glitch, the braking index, n, can be found from ${\rm F}\ddot{{\rm
F}}\,\dot{\rm F}^{-2}$.
We find a braking index of $1.6\pm0.1$ for the data 150 days after the
glitch. This is to be compared with the value Lyne \etal\ obtained by
the same method of $1.5\pm0.4$. Lyne \etal\ improved on this value by
extrapolating $\dot{{\rm F}}$ back to the epoch of the glitch.  We
will report the full analysis, including this extrapolation, in a
future paper.

The DM, which we can deduce from the multiple frequencies observed,
continues to fall as reported in Hamilton \etal\ (1985).
At the epoch of 53193 it was 67.74\,pc\,cm$^{-3}$ and is falling by
4.3\,pc\,cm$^{-3}$ per century. 

We don't see the strong correlation between glitch size and time
between glitches in our data, as reported for PSR0537-6910
(Middleditch \etal\ 2006) 
recently. The best linear fit through the origin gives 46.5 days
$\mu$Hz$^{-1}$.
The data are shown in Figure 4. 

\section{Conclusions}
\label{sec:con}

The Vela pulsar has been timed for more than twenty years, and
continues to provide new insights into the pulsar EOS by, for example,
providing very low limits for the spin-up time and therefore the crust
thickness.  There is clearly a need to observe a glitch with higher
sensitivity and time resolution to investigate both the fast decay
term and to detect the spin-up. Both of these values relate directly
to the pulsar EOS and will provide rich fodder for theoretical
analysis by allowing the measurement of the moment of inertia of the
crust. These observations could be made with the 14-metre telescope
and the enhanced back-end if dedicated observing is continued. If the
observation program is terminated then this effect is a ripe target
for the next generation of pulsar telescopes that could monitor a large
number of targets simultaneously with new beam forming techniques.

\begin{figure}
\centering
  \includegraphics[angle=0,width=0.5\textwidth]{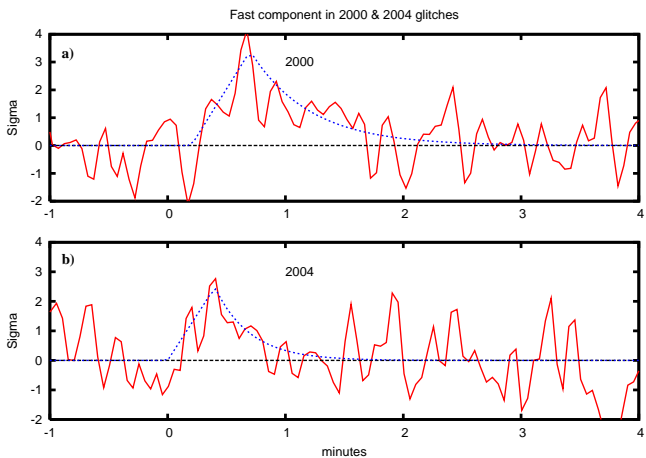}
\caption{The timing residuals after removing the long scale timing
  models, plotted as standard deviations, against the minutes relative
  to the assumed glitch epoch. The model of a later glitch epoch, and
  a very fast spin down, are overlaid. 1a is for 2000, 1b is for 2004.}
\label{fig:fast}       
\end{figure}
%
\begin{figure}
\centering
  \includegraphics[angle=0,width=0.5\textwidth]{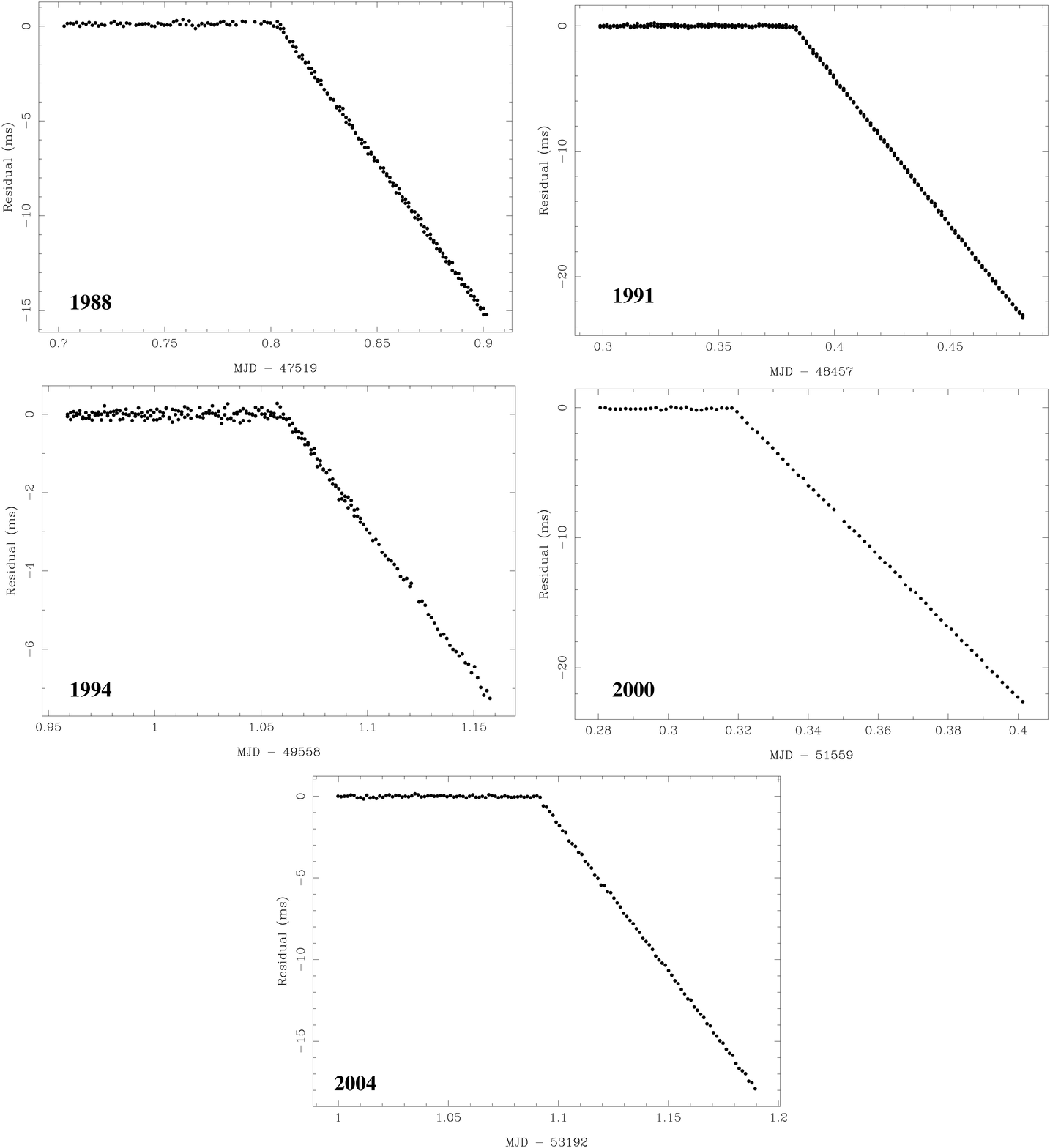}
\caption{Plots of all the glitches of Vela directly observed. The
  residuals are in milliseconds are plotted against time in days.}
\label{fig:glitches}       
\end{figure}
%
\begin{figure*}
\centering
  \includegraphics[angle=270,width=0.7\textwidth]{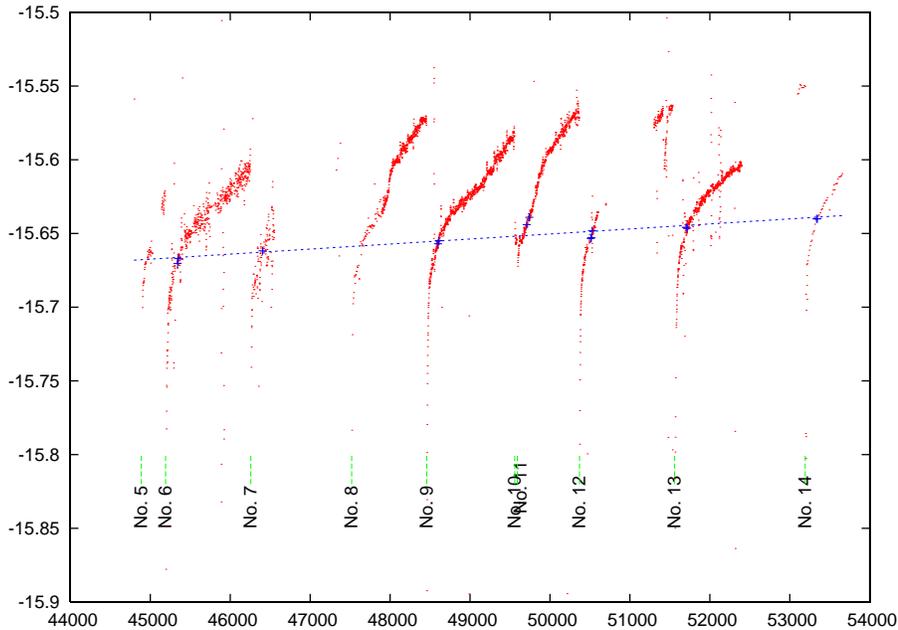}
\label{fig:fdot}       
\begin{center}
\caption{Solutions for $\dot{{\rm F}}$ over twenty years, with the 
$\dot{{\rm F_{150}}}$ marked and the fitted slope for $\ddot{{\rm F}}$
  overlaid.}
\end{center}
\end{figure*}
%

\begin{figure}
\centering
  \includegraphics[angle=0,width=0.5\textwidth]{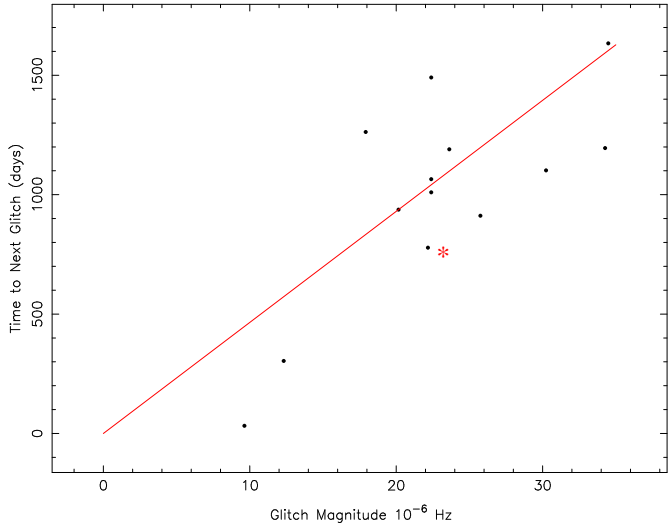}
\label{fig:gm_tng}       
\caption{Plot of the time to the next glitch vs glitch amplitude
  (c.f. Middleditch \etal\ 2006 figure 7), showing the weak correlation
  for the Vela pulsar compared to that of PSR 0537-6910. Marked with a
  red star, but not included in the fit, is the glitch of August 2006
  (Flanagan \& Buchner 2006).}
\end{figure}
\begin{table}[t]
\caption{Parameters for the glitch of MJD 53193.092. The data fit is
  from MJD 53171 through to 53264 (mid-June to mid-September
  2004). All significant figures are given. The model fitted consists of a
  permanent change in the rotation frequency and deceleration (denoted
  with a $p$ subscript) and a number of temporary changes in rotation
  frequency (denoted with a $n$ subscript) which decay on a timescale
  of $\tau$.}
\begin{center}
\begin{tabular}{|c|c|c|}
\hline
\multicolumn{3}{|c|}{Parameters with reference to Epoch MJD 53193}\\
\hline
${\rm F}/Hz$ & $\dot{\rm F}/Hz~s^{-1}$ & $\ddot{\rm F}/Hz~s^{-2}$ \\
\hline
11.1924472071183043 & -1.555028E-11 & 5.27E-23 \\
\hline
\end{tabular}

\begin{tabular}{|l|l|}
$\Delta {\rm F}_p/Hz$ & $\Delta \dot{\rm F}_p/Hz~s^{-1}$ \\
\hline
2.2865E-05&-1.0326E-13\\
\hline
$\tau_n$ & $\Delta {\rm F}_n/10^{-6} Hz$\\
\hline
$1\pm0.2$mins & 54\\ 
00.23 days& 0.21\\ 
02.10 days& 0.13\\ 
26.14 days& 0.16\\ 
\hline
DM & 67.74\,pc\,cm$^{-3}$ \\
\hline
\end{tabular}
\end{center}
\label{tab:1}       
\end{table}




%
%


\end{document}